\documentclass[structabstract]{aa}  
\usepackage{graphicx}
\usepackage{txfonts}
\begin{document}
   \title{Rotationally Resolved Spectroscopy of (20000) Varuna in the near-Infrared}

   \author{V. Lorenzi\inst{1}
          \and
          N. Pinilla-Alonso\inst{2}
          \and
          J. Licandro\inst{3}
          \and
          C.M. Dalle Ore\inst{4}
          \and
          J.P. Emery\inst{2}
   \institute{Fundaci\'on Galileo Galilei-INAF, Spain
              \email{lorenzi@tng.iac.es}
          \and
             University of Tennessee, TN, USA
          \and
             Instituto Astrof\'isico de Canarias, IAC, Spain
          \and
             Carl Sagan Center, SETI Institute and NASA Ames Research Center, CA, USA}          
             }
         
   \date{Received; accepted}

 
  \abstract
   {Models of the escape and retention of volatiles by minor icy objects exclude the presence of volatile ices on the surface of 
   trans-Neptunian objects (TNOs) smaller than $\sim1000$ km in diameter at the typical temperature in this region of the Solar System.
   Whereas, the same models show that water ice is stable on the surface of objects over a wide range of diameters.
   Collisions and cometary activity have been used to explain the process of surface refreshing of TNOs and Centaurs. 
   These processes can produce surface heterogeneity that can be studied by collecting information at different rotational phases.}
   {The aim of this work is to study the surface composition of (20000) Varuna, a TNO with a diameter $668^{+154}_{-86}$km, 
   and to search for indications
    of rotational variability.}
   {We observed (20000) Varuna during two consecutive nights on January 2011, with the near-infrared camera and spectrometer NICS 
   at the Telescopio Nazionale Galileo, La Palma, Spain.
   We used the low resolution mode, with the AMICI prism, obtaining a set of spectra covering the whole rotation period of the Varuna ($Pr = 6.34\ hr$).
   We fit the resulting relative reflectance with radiative transfer models of the surface of atmosphereless bodies.}
   {Studying the spectra corresponding to different rotational phases of Varuna, we did not find any indication of surface variability at $2\sigma$ level. 
   In all the spectra, we detect an absorption at 2.0 $\mu m$, suggesting the presence of water ice on the surface. 
   We do not detect the presence of any other volatiles on the surface, although the S/N is not high enough 
   to discard their presence in small quantities.
   Based on scattering models, we present two possible compositions compatible with our set of data and discuss 
   their implications in the frame of the collisional history of the trans-Neptunian Belt.}
   {We find that the most probable composition for the surface of Varuna is a mixture of amorphous silicates, complex organics and water ice. 
   This composition is compatible with all the materials being primordial, so no replenishment 
   mechanism is needed in the equation. 
   However, our data can also be fitted by models containing up to a 10\% of methane ice.
   For an object with the characteristics of Varuna, this volatile could not be primordial, so an event, such as an 
   energetic impact, would be needed to explain its presence on the surface.}
   \keywords{ Kuiper belt objects: individual: (20000) Varuna -- methods: observational -- methods: numerical -- techniques: spectroscopic -- planets and satellites: composition}
   \maketitle
%

\section{Introduction}
 
The trans-Neptunian objects (TNOs) are thought to be composed mainly of the less processed material from the proto-solar nebula.
The study of their surface composition and their dynamical and physical properties provides important information to understand 
the conditions of the early solar system environment.
Investigation of visible and near-infrared spectra of TNOs and Centaurs (a population of icy objects related to the TNOs and the Jupiter family comets)
show a wide range in compositions. 
Some objects are covered by dark irradiation mantles showing a mostly featureless and neutral 
to reddish spectrum in the visible and near-infrared.
Some other objects have retained large amounts of ices (e.g. $CH_4$, $N_2$, $H_2O$) on their surfaces. 
The spectra of these objects are 
characterized by clear absorption bands 
that appear over the entire wavelength range from the visible to the NIR. 
However, most of the objects show intermediate surface compositions, i.e., a mixture of ices and irradiated mantles. 
Most of these objects are not able to retain volatiles, but water ice can be present on their surface.

Barucci et al. (\cite{bar11}) and Brown et al. (\cite{brown3}) study the possibility of water ice on the surfaces of TNOs and Centaurs  
from extensive collections of spectra of these minor bodies. Both works are in agreement on the broad results.
Most of the objects brighter than an absolute magnitude of H = 3 have deep water ice absorption. 
For those fainter than an absolute magnitude of H = 5, deep water ice absorption is never seen. 
For the intermediate objects, those with a diameter $\sim$ 650 km, only one (2003 AZ$_{84}$) shows a deep water ice absorption band. 

TNOs and Centaurs with a clear detection of water ice are found on all the taxonomic groups, except for the IR 
(objects with a red intermediate slope, Barucci et al., \cite{bar08}).
Moreover, they are found on all the dynamical classes, except the cold classical population (Barucci et al., \cite{bar11})

According to the Minor Planet Center (MPC), (20000) Varuna (hereafter Varuna) is a classical TNO ($a,e,i = 43.189\ AU,0.053,17.1^{\circ}$) with a magnitude $H= 3.6$. 
It has an estimated diameter of $668^{+154}_{-86}$km (Lellouch et al., \cite{lellouch}). 
Its surface composition has been studied through NIR spectroscopy by Licandro et al. (\cite{lic1}) and Barkume et al. (\cite{bark}). 
These authors, however, reach different conclusions. Licandro et al. (\cite{lic1}) find a hint of an absorption of water ice around 2.0 $\mu m$ 
on a reddish NIR spectrum, while Barkume et al. (\cite{bark}) show a featureless spectrum with a blue slope at the H band. 
They interpret these contradictory results as an indication of surface heterogeneity.

In this work, we present the results of a observational campaign designed to study the surface composition of Varuna 
at different rotational phases. 
The purpose of this work is to investigate the possibility of inhomogeneity suggested in previous studies.

\section{Observations and Data}

In January 2011 we observed Varuna during two consecutive nights using the $3.58\ m$ Telescopio Nazionale Galileo (TNG), 
situated at the ``Roque de los Muchachos'' observatory in La Palma, Canary Islands, Spain. 
We used the high-throughput low resolution mode of the Near-Infrared Camera and Spectrometer (NICS), 
with an Amici prism disperser and the $1''$ slit. This mode yields a $0.8-2.5\ \mu m$ spectrum 
with a constant resolving power of $R \simeq 50$. 
The slit was oriented at the parallactic angle (and updated during the night), 
and the tracking of the telescope was set to compensate the non-sideral motion of the TNO.

We observed the object each night during several hours, in order to cover entirely its rotation period $Pr = 6.34\ hr$ (Thirouin et al., \cite{thirouin}). 
The series of spectra consisted of consecutive exposures of $90\ s$ following an ABBA scheme, 
where A is the position of the object in the slit during the first acquisition and B is a position shifted of $10''$ along the slit. 
The ABBA scheme was repeated several times in order to increase the signal to noise ratio (S/N) and to cover the whole rotation period. 
To study possible spectral variations with the rotational phase,  we divided the AB pairs in four groups, each group covering a quarter of the rotation. 
Then we combined each AB pairs to remove the sky contribution from the spectra.
After discarding those spectra with low S/N (probably due to non-photometric conditions), 
we kept 114 spectra for Phase 1, 36 for Phase 2, 48 for Phase 3 and 68 for Phase 4, 
for total exposure times of 10260, 3240, 4320 and 6120 seconds respectively (see Tab. \ref{obs}). 
We combined all the spectra of each group, obtaining a spectrum for each phase that we extracted and calibrated using IRAF standard packages
as described in Licandro et al. (\cite{lic1}).

 \begin{table}
         
          \begin{tabular}{lllc}
            \hline
            \hline
             \textbf{Rot. \  Phase} & \textbf{Date} & \textbf{Time (U.T.)} &  \textbf{Exp.\ Time}\\
             \hline
            Phase\  1& 2011/01/08&21.99-23.51 &  2700 s \\ 
                     &2011/01/09 & 4.35-5.85 &  3960 s \\ 
                    & 2011/01/09 & 23.53-0.93 &  3600 s \\ 
           Phase\  2 &2011/01/08 &23.67-0.93& 1440 s \\ 
                     &2011/01/09 & 5.96-6.28&   1080 s \\ 
                     &2011/01/10 &0.96-2.51 & 720 s \\ 
            Phase\  3 & 2011/01/09&1.55-2.73 &  900 s\\ 
                     & 2011/01/10& 2.54-3.96&  3420 s \\ 
            Phase\  4 &2011/01/09 & 2.77-4.32 & 3600 s \\ 
                      & 2011/01/09& 23.06-23.38&  1080 s\\ 
                      & 2011/01/10& 4.07-4.92&  1440 s \\ 
            \hline
         \end{tabular}
       \caption{Date and time (U.T.) of the observations and total exposure time corresponding 
to the four rotational phases in which we have divided our Varuna data.}
\label{obs}
        
   \end{table}

In order to correct for telluric absorptions and to obtain the relative reflectance, 
we observed each night three stars chosen from this short list: 
P330E (Colina \& Bohlin, \cite{bib4}), Landolt (SA) 93-101,  Landolt (SA) 102-1081 and Landolt (SA) 107-998 (Landolt, \cite{land}).
All of them are usually used as solar analogs and
they were observed at different times in order to cover an airmass range similar to that covered by the object (1.01-2.07).

\begin{figure}
  \begin{center}
   \includegraphics[width=9cm]{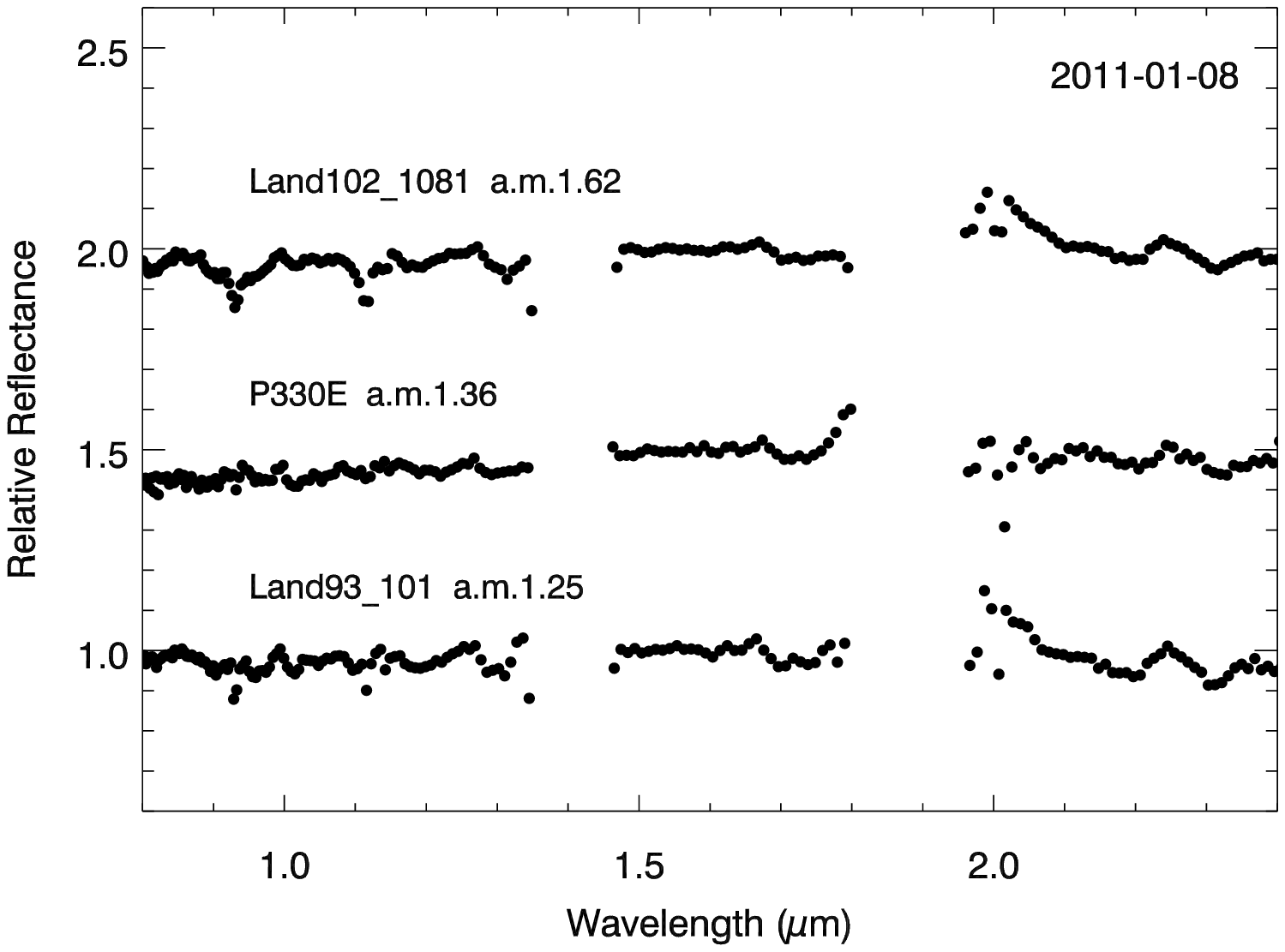}  
   \includegraphics[width=9cm]{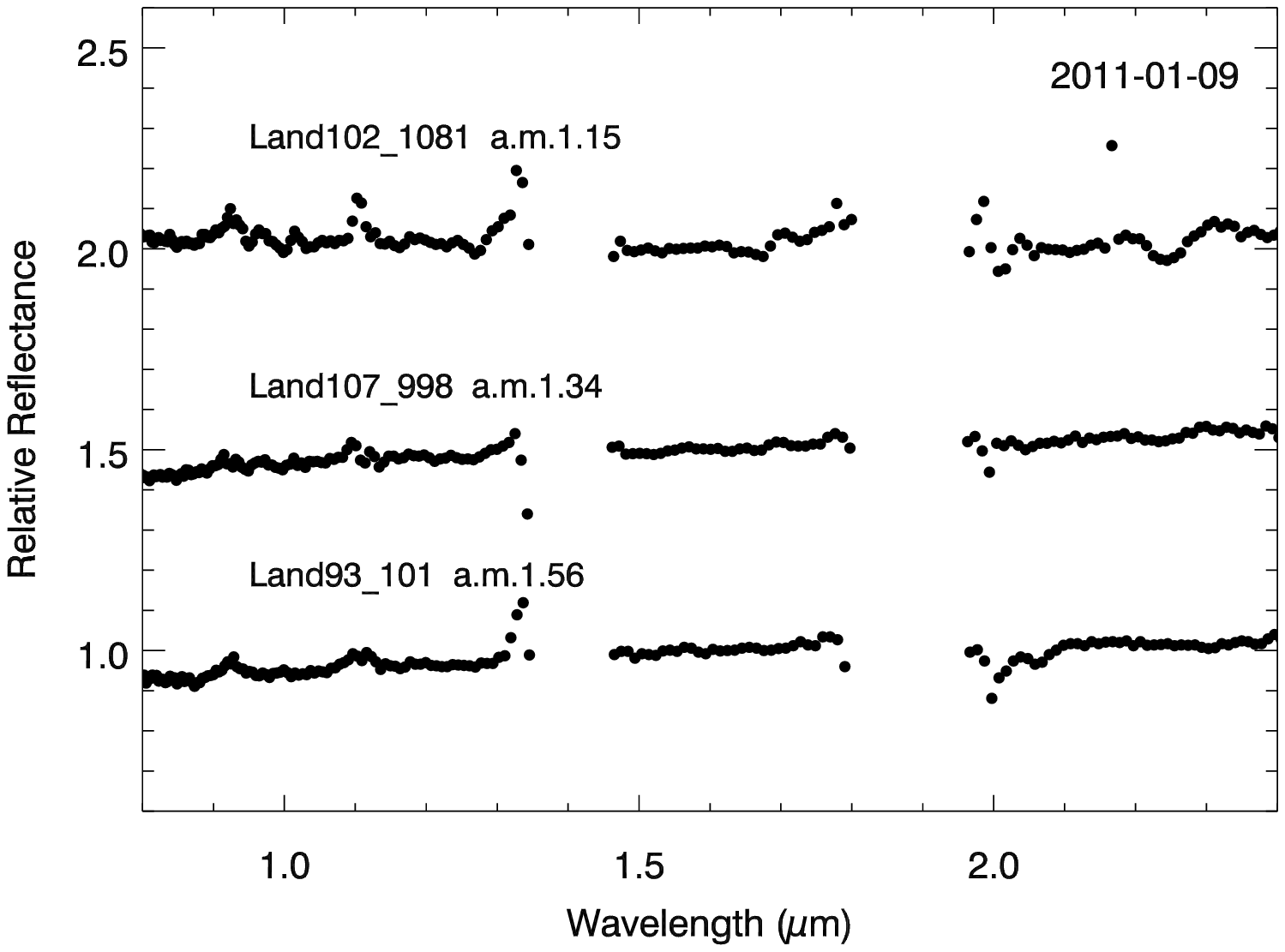}
   \end{center}
      \caption{\textit{Top panel}: Spectra of some pairs of the standard stars observed during the night of 2011-01-08 divided by a pair
      of Landolt (SA) 102-1081 observed at an airmass of 1.25. 
      \textit{Bottom panel}: Same for the night of 2011-01-09 taking a pair of Landolt (SA) 102-1081 at an airmass of 1.41 as a reference.
      The airmass of each standard is included in the legend and the spectra are shifted of $0.5$ in the vertical for clarity.}
         \label{stand}
   \end{figure}

Before comparing the spectra of the target with the spectra of the solar analogs to remove the signature of the Sun, 
we analyzed the spectra of the standard stars with a view to detect small differences in colors,
introduced during the observations, i.e. by inconsistent centering of the star in the slit. These differences could propagate into the spectrum of the target through the reduction process. 
To quantify these errors, we extracted all pairs of stars acquired during the
run. Then we divided, for each night, all of the spectra of the solar analogs by one that we take as reference (Landolt (SA) 102-1081). 
Notice that the resulting of dividing the spectrum of a solar analogs against another should result
in a straight line with spectral slope $S'= 0$. 
An example is shown in Fig. \ref{stand} where small differences in spectral slope can be detected over all the adopted wavelength range.
Some spectral structure appears also close to the wavelength regions affected by telluric absorptions; these are typical of variation of the atmospheric conditions.
Calculating the slope of these ratios and averaging them, we estimate that the systematic errors are no larger than $0.65\%/1000$ \AA{}.

After analyzing the small variation among all the pairs of each standard, we averaged all of them and used the resulting spectrum of each solar analog
to extract the relative reflectance of the target at the four rotational phases. 
We show in the top panel of Fig.~\ref{fases} the four final scaled reflectances  
normalized to unity at $1.6 \ \mu m$ and shifted by $0.5$ in relative reflectance for clarity.
We do not show the values around the two large telluric bands ($1.35-1.46$ and $1.82-1.96 \ \mu m$) 
as the S/N is very low and the difference between the spectrum of the object and those of the solar analogues could introduce false features even in rather stable atmospheric conditions.

\section{Analysis of the spectra}

\begin{figure}
   \centering
   \includegraphics[width=9cm]{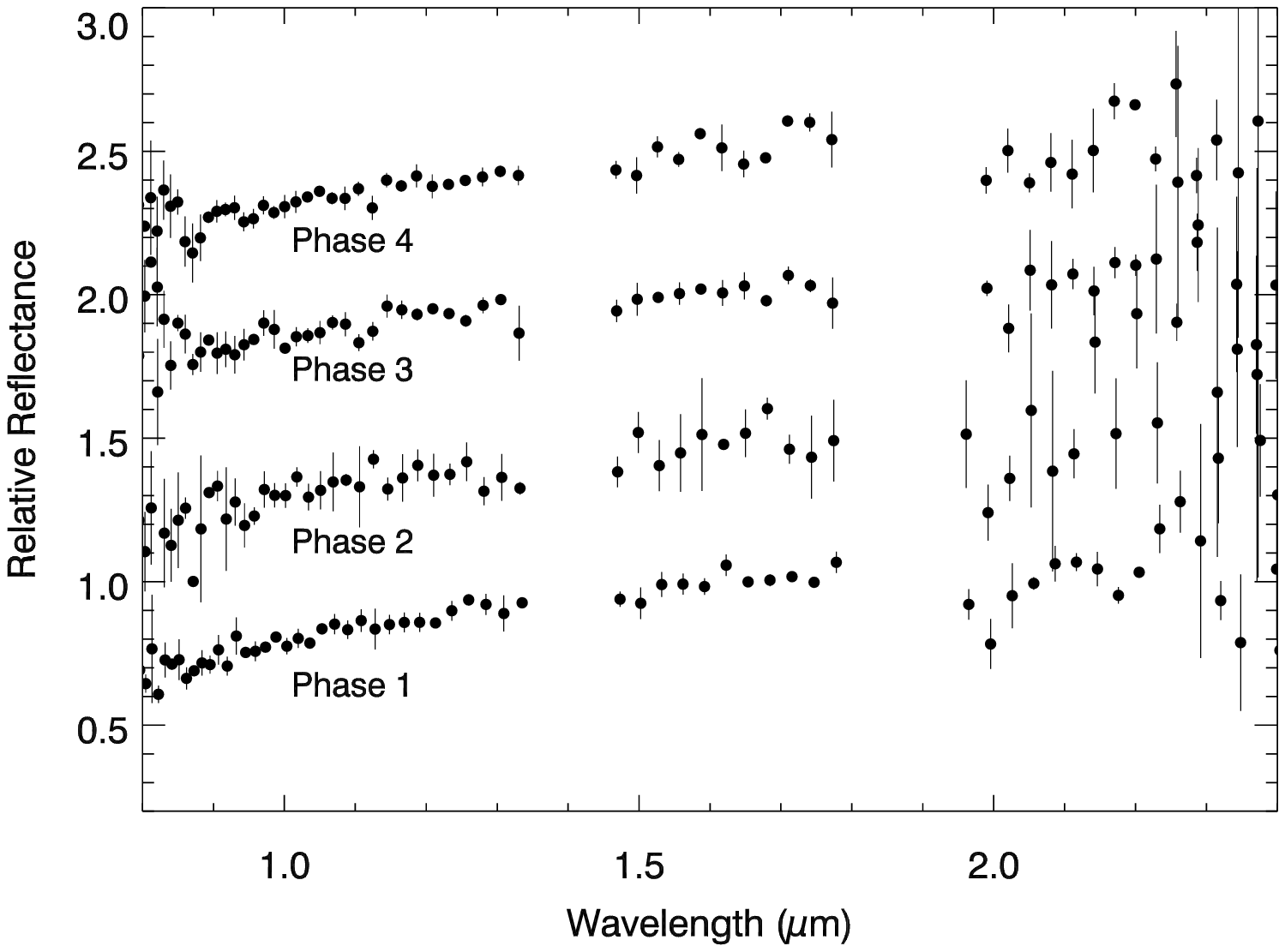}  
   \includegraphics[width=9cm]{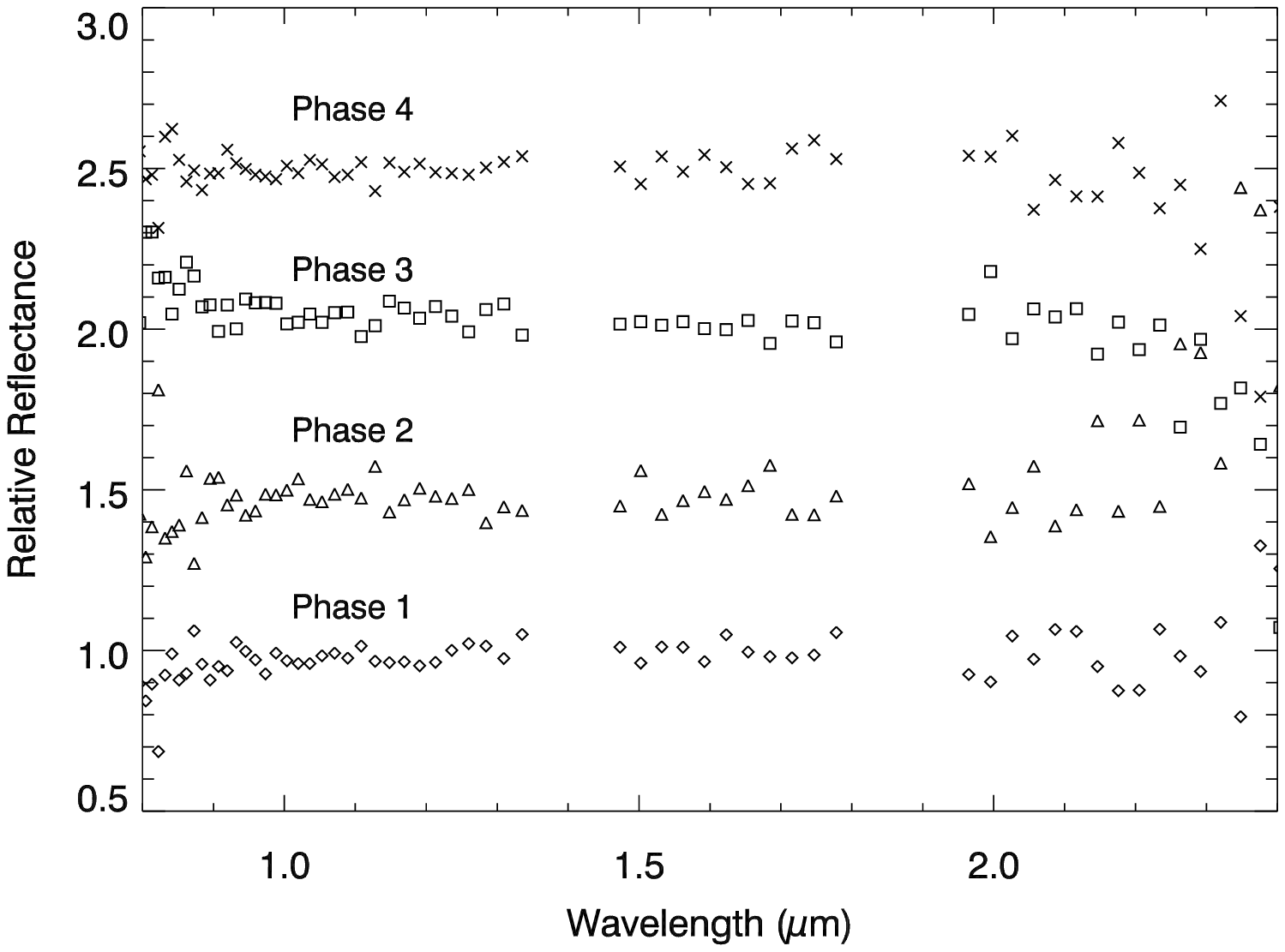}
      \caption{\textit{Top panel}: the four spectra corresponding to the four rotational phases, 
shifted by $0.5$ in relative reflectance for clarity. 
               \textit{Bottom panel:} ratio between the four spectra and the average of all of them, also in this case 
shifted of $0.5$ in relative reflectance. }
         \label{fases}
   \end{figure}

In comparing the four spectra, we first note 
that they are all similar within the S/N (Fig.~\ref{fases} top panel). 
This is more clear in Fig.~\ref{fases} (bottom panel),
where we show the ratio between each of the spectra and the average of all of them, showing all of them a deviation from unity lower than $2\sigma$.
We see that all the spectra are reddish (Fig.~\ref{fases}, top panel) signaling the presence on the surface of Varuna of materials more absorbent at the shorter wavelengths. 
Moreover, the four spectra display an absorption band centered at 2.0 $\mu m$, where the water ice has a deep band.
As suggested in Licandro et al. (\cite{lic1}), the low depth of the 2.0 $\mu m$ band and the non-detection of the water-ice band
at 1.5 $\mu m$ suggest that the fraction of water ice in the surface is not high (we will come back to this in section 4). 
These results are consistent regardless how we define the four phases.

In order to quantify any possible variation in the slope with the rotation phase, 
we compute, for each spectrum, the normalized reflectivity gradient $S'[\% /1000$ \AA{}] in the $0.8-1.8\ \mu m$ range, 
where the continuum can be fitted by a straight line (Jewitt, \cite{jew}). Results are shown in Tab. \ref{results}. 
Note that the errors in this table are computed from the fit, taking into account the dispersion of data in relative reflectance.
However, as mentioned in the previous section, the systematic errors introduced in the slope is not smaller than $0.65\% /1000$ \AA{}.
Hereafter we will use this value for all measurement except those for which the computed errors are larger.
The S' values obtained indicate that the slope does not vary with the rotation within the errors. 

Considering the similarity of the four spectra corresponding to the four phases, 
both in the slope and in the presence of the water ice absorption band,
we have finally combined them to increase the S/N, obtaining an average spectrum for Varuna with a S/N of $\sim35 $ in H band.

\begin{figure}
   \centering
   \includegraphics[width=9cm]{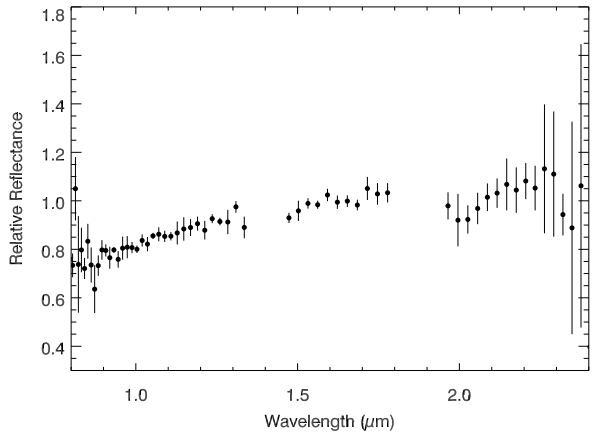}
      \caption{Final averaged spectrum of Varuna normalized at $1.6 \ \mu m$.}
         \label{rot_med}
   \end{figure}

The resulting spectrum of Varuna is shown in Fig. \ref{rot_med}. 
The average spectrum has a normalized reflectivity gradient of $S'=2.94\pm0.65\  \%/1000$ \AA{}  in the $0.8-1.8\ \mu m$ range, 
and a water ice absorption band at 2.0 $\mu m$ with a depth $D= 1- R_b/R_c=21.0 \pm 6.6 \%$, 
where $R_b$ is the reflectance in the center of the band and $R_c$ is the reflectance of the continuum in the same wavelength, 
calculated by fitting a cubic spline on the left and the right sides of the band.

 \begin{table}
         \begin{center}
               \begin{tabular}{lc}
            \hline
            \hline
            \textbf{Rot. Phase} & \textbf{S'[$\%/1000$ \AA{}]}\\
             \hline
            Phase\  1&3.52 $\pm$ 0.16 \\ 
            Phase\  2 &3.11 $\pm$ 0.35 \\ 
            Phase\  3 & 2.95 $\pm$ 0.31 \\ 
            Phase\  4 & 3.47 $\pm$ 0.17 \\ 
            \hline
          
         \end{tabular}
         \end{center}
        \caption{Normalized reflectivity gradient S'(see text for definition) for the spectra corresponding to the four rotational phases.
      The errors are only based on dispersion of the data points, the systematic error is $0.65\% /1000$ \AA{} (see text for details).}
  \label{results}
           
   \end{table}

\begin{figure}
   \centering
   \includegraphics[width=9cm]{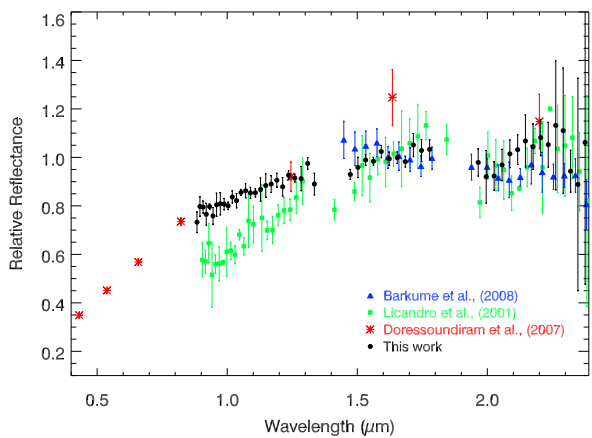}
      \caption{Comparison between our data and data available in the literature.  
All the spectroscopic data are normalized at $1.6 \ \mu m $. The photometric data are merged with the spectra in this work at the J band.}
         \label{comp}
   \end{figure}

If we compare our data with the spectrum from Licandro et al. (\cite{lic1}), we find that the slopes of the two spectra 
are quite different (see Fig. \ref{comp}). 
This is probably due to the fact that Licandro et al. (\cite{lic1}) did not observe a solar analogue star. 
They observed an A0 star for the telluric correction and used the blackbody function of the A0 and G2 star to obtain the relative reflectance. 

The spectrum from Barkume et al. (\cite{bark}), as can be seen in Fig. \ref{comp}, covers only the H and K range. 
It is blue and featureless within the S/N. The discrepancy between that spectrum and the one we show in this work 
cannot be interpreted as an indication of surface heterogeneity, given that 
our spectrum is the average of four almost identical spectra (see Fig. \ref{fases} bottom panel) 
that cover the whole surface of Varuna.

\section{Modelization of the reflectance spectrum}

\subsection{Models with water ice}

From the beginning of the study of the surface composition of Centaurs and TNOs, 
spectra have indicated that they are covered by a mixture of minerals, complex organics and ices (Barucci et al., \cite{bar08}, Dalle Ore et al., \cite{dalleore} and references therein). 
However, it is not trivial to deconvolve a reflectance spectrum into abundances of the different components because the spectra are complex nonlinear functions of grain size, abundance and material opacity (Poulet et al., \cite{poulet}). 
Nevertheless, models of the scattering of light by particulate surfaces have proven to be useful in the study of the surface composition of minor bodies in the Solar System.

In this work we have used the Shkuratov theory (Shkuratov et al., \cite{shku}) to generate a collection of synthetic spectra that reproduce the overall shape of the spectrum of Varuna.
We put special emphasis on the reproduction of the two most representative characteristics of this spectrum, its red slope and the band at 2.02 $\mu$m.
These models use as input the optical constants, the relative abundances and the size of the particles of different materials to compute
the albedo of the surface at each wavelength. 

It is worthwhile to note here that we compare the synthetic spectra with the reflectance of Varuna
using the full visible and near-infrared wavelength range: the spectrum in the NIR presented in this work 
and the photometry in the visible (B, V, R, I bands) (Doressoundiram et al., \cite{dores}).
To give the same weight to the visible part of the data as for the near-infrared when calculating the goodness of the fit, 
we generate a synthetic spectrum in the visible from the photometric data, 
using a spectral resolution of the order of the spectral resolution in the near-infrared and with a SNR determined 
by the error associated with the photometric colors.

As a first step, we run a series of models with different number of components considering materials that are good candidates to be part of 
the surface of minor icy bodies: complex organics (Triton, Titan and Ice tholins), olivines and pyroxenes 
with different content of magnesium and iron, water ice, amorphous carbon
(see Tab. 1 from Emery \& Brown, \cite{emery} for the characteristics of the optical constants of silicates, organics and carbons;
Warren, \cite{war} for the constants of water ice).

We first inspect the fits by eye and find that all the models that better reproduce the spectra include olivine, pyroxene (with a low amount of iron, 20\%), 
triton tholin, amorphous carbon and water ice. 
Based on this approximation we decided to use a five-component model mixing these materials as an intimate mixture (salt-and-pepper).

As a second step, we calculate a grid of models varying the relative abundances of these materials from:
0 to 40\% for water ice, 0 to 50\% for olivine and pyroxene,
0 to 30\% for triton tholin, and 0 to 70\% for amorphous carbon, with a step of 10\% for all of them.
We also vary the size of the particles around the best values suggested by the models on the first step:
5 to 23 $\mu$m, with a step of 3, for water ice; 50 to 200 $\mu$m, with a step of 50 $\mu$m, for the olivine and pyroxene, and
3 to 15 $\mu$m,  with a step of 3, for the tholins.
We use a fixed size of 100 $\mu$m for the carbon, as the dependance of the reflectance of these dark materials 
with the particle diameter is insignificant (Figure 4 from Emery \& Brown, \cite{emery}).
The details of the optical constants used in this model are included in Tab. \ref{materials}.

  \begin{table}
          \begin{tabular}{lcl}
            \hline
            \hline
            \noalign{\smallskip}
            \textbf{Material} & \textbf{Code in ref.}& \textbf{Reference}\\
            \hline
            Amorphous\ water\ ice& & Warren\ (\cite{war})\\
            Amorphous\ olivine&&\\
            $Mg_{2y}Fe_{2-2y}SiO_{3}$ (y=0.5)&O1&Dorschner\ et\ al.\ (\cite{dor})\\
            Amorphous\ pyroxene& & \\
            $Mg_{x}Fe_{1-x}SiO_{3}$ (x=0.8)& P4 &Dorschner\ et\ al.\ (\cite{dor})\\
            Triton\ tholin& &McDonald\ et\ al.\ (\cite{mcd})\\
            Amorphous\ carbon& AC1& Rouleau\ \&\ Martin\ (\cite{rou})\\
            \hline
            \end{tabular}
      \caption{Details of the materials used in the best-fit model with water ice.}
 
         \label{materials}
    
   \end{table}

Finally, we rank the models using the $\chi^{2}$ value, always using the estimated value for the albedo in the visible as a constraint 
(p$_{v}=12.7^{+4.0}_{-4.2}\% $, Lellouch et al., \cite{lellouch}). 

\begin{figure}
   \centering
   \includegraphics[width=9cm]{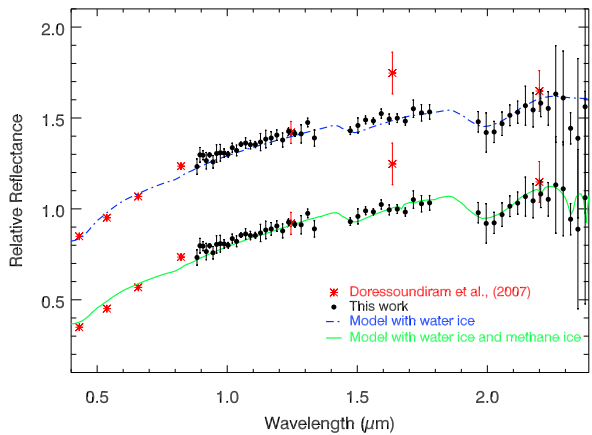}
      \caption{Final spectrum of Varuna together with two models. 
      These models include water ice, silicates, organics and amorphous carbon (top) and water and methane ice, silicates, organics and amorphous carbon (bottom).}
         \label{modelo}
   \end{figure}

The final-best-model consists of an intimate mixture of 25\% water ice (17 $\mu$m), 15\% olivine (50 $\mu$m), 
10\% pyroxene (50 $\mu$m), 35\% tholin (6 $\mu$m) 
and 15\% carbon.
The albedo of this model is 10.7\%, compatible with value estimated in Lellouch et al. (\cite{lellouch}).
The amount of water ice resulting from our theoretical model (25\%) is consistent with previous results (Licandro et al., \cite{lic1}) 
that suggested that water ice could be present on the surface of Varuna, even if the amount might be not very abundant.
The model and spectrum are shown in Fig. \ref{modelo}.

This composition is typical of objects covered by a mantle formed by a mixture of silicates, processed materials (complex organics) and water ice.
Water ice lifetime on the surface of these Centaurs and TNOs is comparable to the age of the solar system, so it could be primordial. No refreshing 
mechanism is needed in this case to explain the surface composition of this Varuna. 

\subsection{Models with methane ice}

Some of the large TNOs (Makemake, Licandro et al. \cite{licmak}; Eris, Brown et al. \cite{broeris}, Licandro et al. \cite{liceris}; 
Pluto, Cruikshank et al. \cite{Cruplu}) are covered by large amounts of methane and other volatiles. The molecule of methane ice 
is optically very active so, although the surface of these bodies may not be dominated by this material, their visible and 
near-infrared spectra are dominated by absorption bands caused by the molecules of methane ice.
Other medium-size TNOs have surfaces that contain or could contain a certain amount of light hydrocarbons such as methane 
ice (e.g. Quaoar, Schaller et al. \cite{schbroqua} and 2007 OR$_{10}$, Brown et al. \cite{brownor10}). However this ice is not expected to be 
common on the surface of most of the TNOs as their temperatures are sufficiently
high and masses are sufficiently low that all the volatiles would have been removed from the surface over solar system 
timescales (Schaller \& Brown, \cite{schbrovol} and Brown et al. \cite{brownrew})

Because of its mass and distance from the Sun, Varuna is expected to have already lost all its original inventory of methane ice from 
the surface. However some fresh methane could have been exposed to the surface by collisions strong enough to break the mantle 
and expose fresh material from the interior of the body. 
This mechanism has been suggested as a mechanism to refresh the surface of TNOs with fresh materials from the interior of the bodies.

To investigate if methane ice could be present on the surface of Varuna we run a final test. We turn our final 5-component model 
into a 6-component model, slightly changing the relative abundances to accommodate some amount of methane. 
We use the optical constants of methane ice at 40 K from Grundy et al., \cite{grund}.
The fact that we do not see any of the stronger absorption bands in the wavelength range covered by our data (e.g. the 2$\nu_{3}$ transition at 1.67 $\mu m$; the 
$\nu_{2}+\nu_{3}+\nu_{2}$ at 1.72 $\mu m$, the $\nu_{3}+$2$\nu_{4}$ at 1.8 $\mu m$ or the $\nu_{2}+\nu_{3}$ at 2.21 $\mu m$) 
puts a strong constraint on the abundance and particle size of the methane in our model. Furthermore, small amounts of methane considerably 
increase the albedo, which reduces the number of possible solutions, given the albedo constraint described above.

We find that the addition of small abundances of methane ice improves (but not strongly affects) the chi-square of our fit. 
Models could contain up to 10\% of methane ice on the surface with a particle size of 20 $\mu m$ by decreasing the amount of water ice to 20\% (20 $\mu m$)
and the amount of amorphous carbon down to 10 \%, giving a surface albedo in the visible of 11.3 \% (Figure. \ref{modelo}).
Our best result with methane ice is not unique (spectral models of mineral/ice mixtures are rarely unique, (Emery \& Brown, \cite{emery}) 
but in this case we have some evident restrictions that limit valid ranges for the relative abundances and particle sizes of the different 
materials. Larger amounts of methane, and/or larger particles make the bands (specially those at 1.72 and 2.21 $\mu m$) more evident so that
they would be detectable in our data. Smaller amounts of silicates or complex organics change the shape of the continuum,  
and models are also strongly constrained by the depth of the band of water ice at 2.02 $\mu m$. 
Larger amounts of high albedo materials (water ice + methane) would immediately increase the albedo in the visible of model above the 
valid value.

But how could we explain the presence of methane ice on the surface of an object like Varuna that should be depleted of volatiles?

One explanation for any possible CH$_4$ would be is that Varuna suffered from a collision in the past. 
The mean probability that a object like Varuna suffers from a collision with other TNOs is $<P_{i}>=1.60\times10^{-22} Km^{-2}yr^{-1}$, 
with medium impact velocity of $<|U|>=2.24\pm1.08\ Kms^{-1}$. This probabilty is even larger if we consider collisions between classical TNOs (Dell'Oro et al., \cite{delloro}).
These kind of energetic impacts can bring fresh ices to the surface 
of a TNO (Gil-Hutton \& Brunini, \cite{gilbru1999}), as less energetic impacts only erode the irradiation mantle. 
As a result of this collision, the irradiation mantle was eroded and some 
ice from the interior (methane and water ice) was sublimated and globally redeposited on the surface. The timescale to distribute vapor globally is of tens of hours,
(Stern, \cite{stern}), which is longer than the rotational period of Varuna, so the material was globally redeposited over the all surface, offering a possible explanation for the homogeneity. The presence of water and methane ices on 
the surface however is not large enough to mask the irradiation mantle, as it has been suggested for other TNOs with a higher content of water ice on 
the surface e.g.1999 TO$_{66}$ (Brown et al., \cite{broTO}) and 2002 TX$_{300}$ (Licandro et al., \cite{licTX}). According to Gil-Hutton et 
al. (\cite{gilEL}) thin layers of ices ($\sim$ 10 - 100 $\mu m$) are not able to mask an irradiation mantle below them.

\section{Conclusions}

We observed Varuna during two consecutive nights, covering twice the rotational period ($\sim 6.34\ hr$).
We show four averaged relative reflectance spectra (0.8 - 2.4 $\mu m$), each of them corresponding to a fourth of the rotation.
Comparing their features (slope and absorption bands) we do not find any indication of surface heterogeneity 
within the signal to noise of our data (larger than the S/N of previously published spectra). 
All our spectra show a reddish slope with an $S'$ ranging from $2.95 \pm 0.31$ to $3.52 \pm 0.16 \% /1000$ \AA{}
and an absorption $\sim2$ $\mu m$ that is indicative of water ice. 
This absorption was previously detected by Licandro et al. (\cite{lic1}) in a single spectrum of Varuna.
The depth and shape of this band is consistent in the four spectra, its medium depth being $21.0 \pm 6.6 \%$. 
In this case, no refreshing mechanism is needed to explain the surface composition of Varuna.

Our fits of the spectral reflectance using models for the scattering of light indicate that Varuna has a processed surface with some ice content,
showing that highly processed materials (complex organics) and silicates coexist with water ice. 
This composition is homogeneous over all the surface of the body at the spatial scale covered by our data.
Water ice lifetime on the surface of these bodies is comparable to the age of the object, so the ice could be primordial.

Our data do not show any indication of other volatiles (such as methane ice) on the surface, although the S/N is not high enough 
to discard their presence. In fact, models show that 10\% of methane ice with a particle size of
20 $\mu m$ improves the fit while still keeping the value of the albedo compatible with value estimated by Lellouch et al. (\cite{lellouch}).
The presence of this ice would indicate that Varuna had suffered a recent energetic impact that would be the responsible of the break up of the mantle
of silicates and organics and of the replenishment of the surface with fresh material from the interior. This kind of impact also results
in homogeneous surfaces as the one we observe for Varuna.

More observations in the H and K band, specifically designed to detect methane ice would be optimum to disentangle the various scenarios
for the recent history of Varuna.

\begin{acknowledgements}
N.P.A. acknowledges support from the project AYA2011-30106-C02-01 and the Juan de la Cierva Fellowship Programme of MINECO (Spanish Ministry of Economy
and Competitiveness). J.L. acknowledges support from the project AYA2012-39115-C03-03 (MINECO). 
Based on observations made with the Italian Telescopio Nazionale Galileo (TNG) 
operated on the island of La Palma by the Fundaci\'on Galileo Galilei of the INAF (Istituto Nazionale di Astrofisica) 
at the Spanish Observatorio del Roque de los Muchachos of the Instituto de Astrof\'isica de Canarias.
We want to thank the referee F. de Meo for her valuable comments that improved the manuscript.
\end{acknowledgements}

\end{document}